\documentstyle[12pt,doublespace,psfig]{article}

\def\ppap{
\mathrel{\hbox{\rlap{\hbox{\lower4pt\hbox{$\sim$}}}\hbox{$<$}}}}
\def\qqap{
\mathrel{\hbox{\rlap{\hbox{\lower4pt\hbox{$\sim$}}}\hbox{$>$}}}}

\newcounter{ref}%
\newcommand{\bib}{\refstepcounter{ref}{\scriptsize\arabic{ref}}}
\newcommand{\wbib}[1]{{\ref{#1}.}~}
\def\up#1{\hbox{${}^{\hbox{\sc #1}}$}}

\def\v@lp@ge:#1,#2:{{\bf#1},~#2}
\def\volumeandpage#1{\v@lp@ge:#1:}

\vspace*{-1cm}

\begin{document}

\centerline{\Large \bf Selective Absorption Processes as the}
\centerline{\Large \bf Origin of Puzzling Spectral Line Polarization from the Sun}
\vspace*{1cm}
\centerline{J.~Trujillo~Bueno$^{*,\dag}$, E.~Landi Degl'Innocenti$^+$, 
M. Collados$^{*}$, L.~Merenda$^{*,{\dag\dag}}$ \& R.~Manso~Sainz$^{+,*}$}
\vspace*{1cm}
\centerline{$^{*}$ Instituto de Astrof\'\i{}sica de Canarias,
             E-38200 La Laguna, Tenerife, Spain.}
\centerline{$^{+}$ Dipartimento di Astronomia e Scienza dello Spazio,
            Universit\`a di Firenze, Italy}	    
\centerline{$^{{\dag\dag}}$Dipartimento di Fisica, Universit\`a di Trento, Italy}	    	     
\centerline{$^\dag$ Consejo Superior de Investigaciones Cient\'\i ficas, E-28006 Madrid, Spain.} 
\vspace*{1cm}
\centerline{{\sl e-mail address:} \ \ jtb@ll.iac.es}

\vfil
\centerline{Published in {\bf NATURE} (24 January 2002)}

\vfil\eject
\noindent

{\bf Magnetic fields play a key role in most 
astrophysical systems, from the Sun to active galactic 
nuclei\up{\bib\label{parker79},\bib\label{carolus00},\bib\label{bland01}}.
They can be studied through their effects on atomic energy levels,
which produce polarized spectral lines\up{\bib\label{sten94},\bib\label{landi01}}. 
In particular,
anisotropic radiation pumping processes\up{\bib\label{cohen66},\bib\label{happer}}
(which send electrons
to higher atomic levels) induce population imbalances that are modified
by weak magnetic fields\up{\bib\label{hanle24},\bib\label{tru01}}. 
Here we report peculiarly polarized light
in the He {\sc i} 10,830-\AA\ multiplet observed in a coronal filament
located at the centre of the solar disk. We show that the polarized
light arises from selective absorption from the ground level
of the triplet system of helium, and that it implies the presence
of magnetic fields of the order of a few gauss
that are highly inclined with respect to the solar radius vector.
This disproves the common 
belief\up{\scriptsize\ref{sten94},\bib\label{sten97},\bib\label{sten01}} 
that population imbalances in long-lived
atomic levels are insignificant in the presence of inclined fields 
with strengths in the gauss range, and demonstrates the
operation of the ground-level Hanle effect in an astrophysical
plasma.}

The Zeeman effect and optical pumping 
are two mechanisms capable of inducing 
polarization signals in the spectral lines that originate
in the outer layers of stellar atmospheres. 

The Zeeman effect\up{\bib\label{consho}} requires the presence
of a magnetic field, which causes the 
atomic energy levels to split into different magnetic
sublevels. This splitting produces local sources and sinks
of light polarization because of the ensuing wavelength
shifts of transitions between levels. The Zeeman effect
is most sensitive in {\it circular} polarization, with a magnitude
that scales with the ratio between the Zeeman splitting
and the width of the spectral line (which is very much
larger than the natural width of the atomic levels).
This so-called {\it longitudinal} Zeeman effect 
responds to the line-of-sight component of the magnetic field.
In contrast, the {\it transverse} Zeeman effect
responds to the component of the magnetic field perpendicular
to the line of sight, but produces {\it linear} polarization signals
that are normally negligible for magnetic strengths ${\rm B}\,{<}\,100$ gauss.

Anisotropic radiation pumping\up{\scriptsize\ref{cohen66},\scriptsize\ref{happer}}
produces {\it atomic level polarization}--that is, 
population imbalances and quantum interferences between the sublevels of
degenerate atomic levels (Fig. 1).
The presence of a magnetic field is {\em not} necessary
for the operation of such optical pumping processes, which can be
particularly efficient in creating atomic polarization if the depolarizing rates  
from elastic collisions are sufficiently low. As clarified below,
the mere presence of atomic polarization of the type illustrated
in Fig. 1 implies local sources and sinks of {\it linear} polarization.
The Hanle effect\up{\scriptsize\ref{hanle24},\scriptsize\ref{tru01}} 
(Fig. 1) modifies the {\em atomic polarization}
of the unmagnetized reference case, and
gives rise to a complicated magnetic field 
dependence of the {\it linear} polarization of the scattered light,
which is being increasingly applied as a diagnostic tool
for weak magnetic fields in astrophysics.

It is often assumed that the observable effects of atomic 
polarization of long-lived atomic levels are strongly suppressed by 
Hanle depolarization in the presence of solar magnetic fields
that are highly inclined with respect to the solar radius and have strengths 
in the gauss range\up{\scriptsize\ref{sten94},\scriptsize\ref{sten97},
\scriptsize\ref{sten01}}. 
Since some of the many ``enigmatic'' signals 
of scattering polarization that have been observed 
within the edge of the solar 
disc\up{\bib\label{sten-keller97},\bib\label{lin-pen98},
\bib\label{sten-keller-gandor00}} 
have been shown to be due to ground state atomic 
polarization\up{\bib\label{lan98},\bib\label{tru99},\scriptsize\ref{tru01}}, 
and because milligauss (or weaker) magnetic fields are believed to be very rare in 
the highly conductive solar plasma, it has been concluded that the field 
must be oriented fairly close to the radial direction wherever the 
signatures of lower level atomic polarization are 
observed\up{{\scriptsize\ref{sten01}}}. Although
this conclusion is probably valid for the observations of scattering
polarization in the Na {\sc i} D$_1$ line\up{\scriptsize\ref{lan98}}, 
it should not be generalized to all
cases in which the signatures of lower-level atomic polarization are observed.
Moreover, the observable effects of lower level atomic polarization, 
whether the Hanle effect destroys or creates linear polarization signals
in spectral lines, depend on the scattering geometry. To clarify 
the situation, it is necessary to investigate carefully to what 
extent observable effects of the atomic polarization of long-lived 
atomic levels can not only survive partial Hanle destruction but even 
be enhanced by horizontal magnetic fields in the gauss range.

Such investigations can be performed in magnetized astrophysical plasmas,
such as those in solar prominences and filaments.
Solar prominences and filaments are relatively cool and dense
ribbons of plasma embedded in the $10^6$ K solar corona.
The ribbons are thought to be
confined by the action of highly inclined
magnetic fields with strengths in the gauss range\up{\bib{\label{balle01}}}.
Prominences and filaments are in fact the same phenomenon
but observed in different circumstances.
Both prominences and filaments absorb the 
photons from the underlying solar photosphere
and re-emit them in all directions. But
prominences are observed outside the visible outer edge of the Sun
(that is, against the dark background of the sky), while filaments are 
the same types of structures observed against the bright background of the solar disk. 
Therefore, we see emission lines in prominences, but absorption lines
in filaments. 

Figure 2 contrasts prominences and filaments and illustrates
our theoretical prediction concerning the expected linear polarization
of a line transition with $J_l=1$ and $J_u=0$,
where $J_l$ and $J_u$ are the angular momentum quantum numbers
of the lower and upper levels, respectively. 
In fact, there are two key mechanisms by means of which
atomic level polarization can generate linear polarization signals
in spectral lines: the first is due to the emission process
(that is, to the atomic polarization of the upper level),
while the second is subtly related with the absorption process
(that is, to the atomic polarization of the lower level). In general,
the first mechanism (caused exclusively by the spontaneous emission
events that follow the anisotropic radiative excitation) is the only one that
is taken into account\up{{\scriptsize\ref{sten94}},\bib{\label{fauro94}},
{\scriptsize\ref{sten97}},{\scriptsize\ref{lin-pen98}},{\scriptsize\ref{sten01}}}.  
The role played by the atomic polarization of the lower level
in producing linear polarization via the 
absorption process\up{\bib\label{trulan97}}
has never been considered seriously.

We have observed the intensity and polarization
of the ${\rm He}$ {\sc i} 10830 \AA$\,$ multiplet
in a variety of solar prominences and filaments
using the Tenerife Infrared Polarimeter\up{\bib{\label{TIP}}}
attached to the Vacuum Tower Telescope\up{\bib{\label{VTT}}}. 
The ${\rm He}$ {\sc i} 10830 \AA$\,$ multiplet originates between a lower term
($2^3S_1$) and an upper term ($2^3P_{2,1,0}$). 
Therefore, it has three spectral lines\up{\bib{\label{radzig}}}: 
a `blue' line at $\lambda 10829.09$
(with $J_l=1$ and $J_{u}=0$), and two `red' lines at $\lambda 10830.25$
(with $J_{u}=1$) and at $\lambda 10830.34$ (with $J_{u}=2$) which
appear blended at the plasma temperatures of prominences and filaments.
As explained in Fig. 2 legend, 
detection of a significant linear polarization signal
in the `blue' line ($J_l=1$ and $J_u=0$) would be due to
the atomic polarization of the lower level with $J_l=1$.
We note that this lower level is {\it metastable}--that is,
it is a relatively long-lived atomic level
whose atomic polarization is vulnerable 
(via the ground level Hanle 
effect\up{\bib\label{landolfi86},\scriptsize\ref{tru01}})
to magnetic fields of very low intensity (${\sim}10^{-3}$ gauss).

We first consider prominences. The data points in Fig. 3
show the four Stokes parameters observed in a prominence.
As expected from the theoretical prediction of Fig. 2, the
`blue' line of the ${\rm He}$ {\sc i} 10830 \AA$\,$ multiplet does {\em not} show any
significant linear polarization, which implies that this particular
prominence has a very small optical thickness along the line of sight.
However, there are very significant Stokes $Q$ and $U$ signals
around the wavelengths of the blended `red' components. These linear
polarization signals are the observational signature of the
atomic polarization of the {\it upper} levels with $J_{u}=2$ and $J_{u}=1$.
Note that there
are sizeable circular polarization signals in both the `blue' and `red' components.
They are the result of the longitudinal Zeeman effect. Their detection 
is essential for the determination of the intensity of the
magnetic field because for fields larger than only a few gauss
the ${\rm He}$ {\sc i} 10830 \AA$\,$ multiplet
enters into the saturated Hanle effect regime for the upper level, where the
linear polarization signals are sensitive only to the
orientation of the magnetic field vector.

The solid line in Fig. 3 gives the result of our theoretical modelling 
taking into account the influence of ground-level polarization.
From the fit to the observation
we infer a magnetic field of about 40 gauss, inclined
at 31$^\circ$ to the radial direction through the observed
point. Interestingly, the dotted line shows what happens if we 
carry out the calculation with this magnetic field vector, but
assuming a completely unpolarized ground level. In the present prominence case,
the significant difference with respect to the solid-line calculation
is solely the result of
the important feedback that the existing ground level polarization 
is producing on the atomic polarization of the upper 
levels with $J_{u}=2$ and $J_{u}=1$.
We note that a magnetic field diagnostic of solar prominences
that neglects the influence of ground-level polarization would imply
a significant error (${\sim}10^\circ$) in the field orientation and
in the magnetic strength (${\sim}$ a few gauss).

We now turn our attention to filaments. The data points in
Fig. 4 show the observed Stokes parameters in  
a solar filament that was situated exactly at the centre of the solar disc
on 2 June 2001. We have selected this coronal filament 
in order to demonstrate that
linear polarization signals can be produced 
{\em even} at the very centre of the solar disc 
where we meet the forward scattering case.
As seen in the figure, the `blue' line 
now shows a very significant linear polarization signal with a {\it negative}
Stokes $Q$ amplitude, which is of the same
order-of-magnitude as the {\it positive} 
Stokes $Q$ amplitude observed in the
`red' blended component. The observed linear polarization signals are entirely
due to the Hanle effect operating at disc centre. This can be
possible only if there exists a magnetic field with a significant
inclination to the radial direction through the observed
point, otherwise the polarization at disc centre would be zero
for reasons of symmetry. The very existence 
of a sizeable Stokes $Q$ signal in the `blue' line
demonstrates that the $^3{\rm S}_1$ ground level is significantly polarized. 

The solid line in Fig. 4 shows the result of our theoretical modelling
of the Hanle effect at disc centre taking into account the
influence of ground level polarization.
From the fit to the observation
we infer a magnetic field of 20 gauss, inclined
by about 105$^\circ$ to the radial direction through the observed
point and with a horizontal component at an angle
of about 10$^\circ$ in the clockwise direction with respect to the axis of the
filament. The agreement with the spectropolarimetric observation
is remarkable. It demonstrates that the ground level Hanle effect
is operating in the outer solar atmosphere, producing very significant
linear polarization signals by selective absorption from the
unevenly populated magnetic sublevels of a long-lived atomic 
level. 
 
Our results demonstrate that the atomic polarization of long-lived levels,
which is induced by optical pumping processes, generates observable
polarization signatures due to the highly inclined magnetic fields with strengths in 
the gauss range that are characteristic of solar prominences.
Instead of destroying the atomic polarization the magnetic 
fields via the Hanle effect produce linear polarization signals that are 
of the same order of magnitude as those caused by the atomic polarization
of the short-lived excited states. Moreover, our results provide
convincing observational evidence of the operation of an atomic
effect that may have diagnostic use
in several astrophysical contexts. It concerns the creation of linear polarization
signals in spectral lines induced by magnetic fields in forward scattering
and by dichroism. These phenomena reveal
unfamiliar aspects of the Hanle effect and open up a new diagnostic window on 
the investigation of the magnetism of the outer 
solar atmosphere (chromosphere, transition region and corona).

{\bf Acknowledgements}
We thank R. Casini (HAO; Boulder) and J. O. Stenflo (ETH; Zurich)
for discussions on quantum electrodynamics
and for helping with the presentation of the Letter.
The German Vacuum Tower Telescope is operated by the Kiepenheuer Institut
at the Observatorio del Teide of the Instituto de Astrof\'\i sica
de Canarias (IAC). The Tenerife Infrared Polarimeter (TIP) has
been developed by the IAC. We also acknowledge the support of
the European Solar Magnetometry Network
and of the Spanish Plan Nacional de Astronom\'\i a y Astrof\'\i sica.

\baselineskip=.99\baselineskip \vfil\eject\noindent

\newpage

\noindent
\hrule

\noindent
{\bf References}

\footnotesize
\noindent
\begin{itemize}

\item[\wbib{parker79}] Parker, E.N.
   {\sl Cosmical Magnetic Fields}.
   (Clarendon Press, Oxford, 1979)
   
\item[\wbib{carolus00}] Schrijver, C.J. \& Zwaan, C.  
    {\sl Solar and Stellar Magnetic Activity}.
    (Cambridge University Press, 2000)   
   
\item[\wbib{bland01}] Blandford, R., et al.  
    Compact Objects and Accretion Disks.
    In {\sl Astrophysical Spectropolarimetry}
    (eds. Trujillo Bueno, J., Moreno-Insertis, F. \& S\'anchez, F.)
    177-223 (Cambridge University Press, 2002) 
    
\item[\wbib{sten94}] Stenflo, J.O.
   {\sl Solar Magnetic Fields: Polarized Radiation Diagnostics}.
   (Kluwer, Dordrecht, 1994) 
   
\item[\wbib{landi01}] Landi Degl'Innocenti, E. 
    The Physics of Polarization.
    In {\sl Astrophysical Spectropolarimetry}
    (eds. Trujillo Bueno, J., Moreno-Insertis, F. \& S\'anchez, F.)
    1-53 (Cambridge University Press, 2002)        

\item[\wbib{cohen66}] Cohen-Tannoudji, C. \& Kastler, A.
   Optical Pumping.
   In {\sl Progress in Optics}
   (ed. Wolf, E.)
   {\bf 5}, 3 -- 81 (1966).
   
\item[\wbib{happer}] Happer, W. 
     Optical Pumping.
    {\sl Rev. Mod. Phys.} {\bf 44}, 169 -- 249 (1972). 
    
\item[\wbib{hanle24}] Hanle, W. 
     \"Uber magnetische Beeinflussung der Polarisation
     der Resonanzfluoreszenz.
    {\sl Z. Phys.} {\bf 30}, 93 -- 105  (1924). 
    
\item[\wbib{tru01}] Trujillo Bueno, J.
   Atomic Polarization and the Hanle Effect.
   In {\sl Advanced Solar Polarimetry: Theory, Observations and Instrumentation} 
   (ed. Sigwarth, M.)
   ASP Conf. Series, 161 -- 195 (2001). 
   
\item[\wbib{sten97}] Stenflo, J.O.   
   Quantum interferences, hyperfine structure, and Raman scattering on the Sun.
   {\sl Astron. Astrophys.} {\bf 324}, 344 -- 356 (1997).              

\item[\wbib{sten01}] Stenflo, J.O.
   Observations of scattering polarization and the diagnostics of solar magnetic fields.
   In {\sl Advanced Solar Polarimetry: Theory, Observations and Instrumentation} 
   (ed. Sigwarth, M.)
   ASP Conf. Series, 97 -- 108 (2001).

\item[\wbib{consho}] Condon, E.U. \& Shortley, G.H.
   {\sl The Theory of Atomic Spectra}.
   (Cambridge University Press, 1979)   
      
\item[\wbib{sten-keller97}] Stenflo, J.O. \& Keller, C.
   The second solar spectrum: a new window for diagnostics of the Sun.
   {\sl Astron. Astrophys.} {\bf 321}, 927 -- 934 (1997).           
   
\item[\wbib{lin-pen98}] Lin, H., Penn, M.J. \& Kuhn, J.R.
    He {\sc i} 10830 Line Polarimetry: A New Tool to Probe
    the Filament Magnetic Fields.
    {\sl Astrophys. J.} {\bf 493}, 978 -- 995 (1998).   
   
\item[\wbib{sten-keller-gandor00}] Stenflo, J.O., Keller, C. \& Gandorfer, A.
   Anomalous polarization effects due to coherent scattering on the sun.
   {\sl Astron. Astrophys.} {\bf 355}, 789 -- 804 (2000).                            
   
\item[\wbib{lan98}] Landi Degl'Innocenti, E. 
     Evidence against turbulent and canopy-like magnetic fields
     in the solar chromosphere.
    {\sl Nature} {\bf 392}, 256 -- 258 (1998).    
    
\item[\wbib{tru99}] Trujillo Bueno, J.
   Towards the Modelling of the Second Solar Spectrum.
   In {\sl Solar Polarization} 
   (eds. Nagendra, K.N. and Stenflo, J.O.).
   Kluwer Academic Publishers, 73 -- 96 (1999).      
   
\item[\wbib{balle01}] van Ballegooijen, A.A.
   Solar Prominence Models.
   In {\sl Encyclopedia of Astronomy and Astrophysics}.
   Nature Publishing Group (2001).      
   
\item[\wbib{fauro94}] Faurobert-Scholl, M.
     Hanle effect with partial frequency redistribution. 
     I. Numerical methods and first applications
     {\sl Astron. Astrophys.} {\bf 246}, 469 -- 480 (1991). 
     
\item[\wbib{trulan97}] Trujillo Bueno, J. \& Landi Degl'Innocenti, E.
     Linear polarization due to lower-level depopulation pumping
     in stellar atmospheres.
     {\sl Astrophys. J.} {\bf 482}, L183 -- L186 (1997).       
    
\item[\wbib{TIP}] Mart\'\i nez Pillet, V., et al.
    LPSP and TIP: Full stokes polarimeters for the Canary Islands observatories.    
    In {\sl High Resolution Solar Physics: Theory, Observations, and Techniques} 
    (eds. Rimmele, T.R., Balasubramaniam, K. S. \& Radick, R. R.)
    ASP Conference Series, {\bf 183}, 264-272 (1999)
    
\item[\wbib{VTT}] Soltau, D.
    Present and Future Solar Observational Facilities
    of the German Vacuum Tower Telescope.
    In {\sl The Role of Fine-Scale Magnetic Fields 
    on the Structure of the Solar Atmosphere}
    (eds. E.H. Schr\"oter, M. V\'azquez \& A.A. Wyller)
    (Cambridge University Press, 362 -- 366, 1987). 
    
\item[\wbib{radzig}] Radzig, A.A., \& Smirnov, B.M.
   {\sl Reference Data on Atoms, Molecules, and Ions}.
   (Springer-Verlag, 1985)
   
\item[\wbib{landolfi86}] Landolfi, M., \& Landi Degl'Innocenti, E.
   Resonance scattering and the diagnostic of very weak magnetic fields in diffuse media.
   {\sl Astron. Astrophys.} {\bf 167}, 200 -- 206 (1986). 

\item[25.$\,$] Landi Degl'Innocenti, E. 
    The determination of vector magnetic fields in prominences
    from the observations of the Stokes profiles in the D$_3$ line of helium.
    {\sl Solar Phys.} {\bf 79}, 291 -- 322 (1982).
    
\item[26.$\,$] Landi Degl'Innocenti, E. 
    Polarization in Spectral Lines: I. A Unifying Theoretical Approach.
    {\sl Solar Phys.} {\bf 85}, 3 -- 31 (1983).    
    
\item[27.$\,$] Bommier, V.
   Quantum theory of the Hanle effect.
   {\sl Astron. Astrophys.} {\bf 87}, 109 -- 120 (1980). 
   
\item[28.$\,$] Chandrasekhar, S.
   {\sl Radiative Transfer}
   (Dover Publications, New York, 1960)     
   
\end{itemize}

\vfill
\eject

%%\baselineskip=.99\baselineskip \vfil\eject\noindent

\normalsize

{\bf Figure 1. Anisotropic radiation pumping and the Hanle effect}

An unpolarized radiation field can induce population imbalances and 
quantum interferences (or coherences)
between the sublevels of degenerate atomic levels (that is, atomic polarization)
if the illumination of the atomic system is anisotropic.
For example, {\em upper-level population pumping} occurs when some {\em upper state}
sublevels have more chances of being populated than others (case $\bf a$).
On the contrary, {\em lower-level depopulation pumping}
occurs when some {\em lower state} sublevels
absorb light more strongly than others (case $\bf b$).
It is also important to note that line transitions between levels
having other total angular momentum values
(e.g., $J_l=J_u=1$ or with $J_l=1$ and $J_u=2$) 
permit the transfer of atomic polarization
between both levels via a process called {\em repopulation pumping}
(e.g. lower-level atomic polarization can result simply
from the spontaneous decay of a {\em polarized} upper level).  
The Hanle effect is the modification of the atomic polarization
of degenerate atomic levels 
caused by the action of a magnetic field
such that its corresponding Zeeman splitting is comparable 
to the inverse lifetime (or natural width) of the 
degenerate atomic level under consideration.
For the Hanle effect to operate, the magnetic field vector ($\bf B$) has to be 
significantly {\em inclined}
with respect to the symmetry axis ($\bf {\Omega}$) of the pumping radiation field.
The formula used to estimate
the {\em maximum} magnetic field intensity $\rm B$ (in gauss)
to which the Hanle effect can be sensitive is
$10^6{\rm B}g\,\approx\,1/t_{\rm life}$,
where $g$ and $t_{\rm life}$ are, respectively, the Land\'e factor 
and the lifetime (in seconds) of the given atomic
level. In a reference frame whose
$z$-axis (that is, the quantization axis of total angular momentum) is parallel
to the direction of the magnetic field vector, the population imbalances turn out
to be insensitive to the magnetic field, while the coherences are reduced
and dephased as the magnetic strength is increased. This so-called
magnetic field reference frame is the one we have chosen here
while visualizing the induced population imbalances for the ``strong field'' case
(that is, the case for which the coherences are negligible). 
We note that the atomic polarization of a given atomic
level sensitively depends on the complexity of the assumed atomic model.

\vfill
\eject

{\bf Figure 2. The solar prominence case versus the solar filament case} 

In a prominence located in the plane of the sky
at a given distance above the visible outer edge of the Sun 
we see the result of $90^{\circ}$ scattering events, while in a filament 
situated exactly at the centre of the solar disc we see the result of
forward-scattering events. This figure illustrates these
two cases by considering observations of a magnetized plasma ribbon
using polarimeters at positions
``$\bf 1$'' (the prominence case) and ``$\bf 2$'' (the filament case).  
The figure refers exclusively to the `blue' line of 
the ${\rm He}$ {\sc i} 10830 \AA\ multiplet,
which is a line transition with $J_l=1$ and $J_u=0$.
Since the upper level cannot carry any atomic polarization,
the spontaneously emitted radiation which follows the anisotropic radiative excitation
is virtually unpolarized. For this reason, the observer at position ``{\bf 1}''
sees that the fluorescently scattered beam is {\em unpolarized}.
However, if the lower level is polarized as indicated in 
the inset, then the {\em transmitted} beam seen by the observer at position ``{\bf 2}''
will have an excess of linear polarization
{\em perpendicular} to the direction of the horizontal magnetic field,
simply because the $\Delta{M}=0$, or $\pi$ transitions, absorb more efficiently
than the $\Delta{M}=\pm{1}$, or $\sigma$ transitions. 
This selective absorption mechanism\up{\scriptsize\ref{trulan97}} 
is called {\it dichroism}
because the plasma is behaving as a dichroic medium
(that is, the absorption coefficient in the line transition
depends on the polarization of the radiation).
We note that {\it repopulation pumping} plays a crucial role
in polarizing the ground level of the triplet system of ${\rm He}$ {\sc i},
since our calculations are based on a realistic multiterm atomic model
and not on the {\em two-level} model atom considered in Figure 1. This is
why the lower-level polarization shown in the inset is
different from Fig. 1b. 
%(Although a {\em two-term}
%model atom provides a reasonably good approximation for modelling the
%He {\sc i} 10830 \AA\ lines, 
%we have nevertheless used a realistic {\em multiterm} model atom, also because
%of our additional interest on 
%the He {\sc i} D$_3$ line whose lower term
%is the upper term of the He {\sc i} 10830 \AA\ multiplet).  
So for $1{\rightarrow}0{\rightarrow}1$ scattering processes
we expect to observe virtually zero linear polarization in 
optically thin prominences,
but a sizeable linear polarization signal in filaments if 
a significant amount of lower-level atomic polarization is present.

\vfill
\eject

{\bf Figure 3. Prominence case: observation versus theory}

Spectropolarimetric observation of a solar prominence (circles) versus
theoretical modelling taking into account the influence of
ground level atomic polarization
(solid line) or neglecting it (dotted line).
Our modelling assumes that the
${\rm He}$ {\sc i} atoms, lying at a given height ($h$) above the
solar photosphere, are illuminated by an unpolarized 
and spectrally flat radiation field.
It is based on the
quantum theory of the generation and transfer  
of polarized radiation\up{\scriptsize{25,26}}, 
which we have applied describing the
${\rm He}$ {\sc i} atoms
in the incomplete Paschen-Back effect regime\up{\scriptsize{27}}.
The Stokes $I$ parameter quantifies the total {\em intensity}
of the observed light, the Stokes $Q$ and $U$ parameters
represent the degree of {\em linear} polarization along
two reference axis that form an angle of $45^\circ$ between them,
while Stokes $V$ quantifies the degree of {\em circular} 
polarization\up{\scriptsize{28}}.  
The observed prominence region had 
a projected height on the plane of the sky of
$20^{''}$ over the visible edge of the solar disc
(that is, $h\,{\approx}\,15000$ km).
The fit to the observations
was done assuming a magnetic field vector with intensity
${\rm B}\,=\,40$ gauss, inclination $\theta_{\rm B}=31^\circ$, and 
azimuth $\chi_{\rm B}=176^\circ$. Note that
$\lambda_0=10829.09$ \AA$\,$ is the line centre wavelength
of the `blue' component of the ${\rm He}$ {\sc i} 10830 \AA$\,$ multiplet.
The positive reference direction for Stokes $Q$ is perpendicular
to the radial direction through the observed point.
The Stokes profiles are normalized to the maximum
line-core intensity of the `red' emission line.
We point out that for this particular geometry of scattering, the
determination of the magnetic field is ambiguous. The alternative
determination 
${\rm B}\,=\,40$ gauss, $\theta'_{\rm B} = 180^{\circ} - \theta_{\rm B} = 149^{\circ}$,
$\chi'_{\rm B} = -\chi_{\rm B} = -176^{\circ}$, gives the
same theoretical curve.

\vfill
\eject

{\bf Figure 4. Filament case: observation versus theory}

Spectropolarimetric observation 
of a solar filament located at the disc-centre (circles) versus 
theoretical modelling
taking into account the influence of
ground-level atomic polarization
(solid line) or neglecting it (dotted line).
Our choice for the positive
reference direction of the observed
Stokes $Q$ parameter is the one which minimizes Stokes $U$.
The solid-line fit 
has been achieved assuming ${\rm B}\,=\,20$ gauss, $\theta_{\rm B}=105^\circ$ and 
a height $h=40^{''}$ (that is, about 30000 km) above the visible solar surface.
The positive
reference direction for the theoretical Stokes $Q$ profile is parallel to the 
projection of the magnetic field vector on the solar surface.
Therefore, negative Stokes $Q$ values indicate that 
the linear polarization
of the observed beam is {\em perpendicular} to the magnetic field of the filament plasma,
as illustrated in Fig. 2. The dotted line neglects the influence
of ground level polarization, but takes into account the
(negligible) contribution of the transverse Zeeman effect.  
The Stokes $I$ profile is normalized to the local continuum intensity,
while Stokes $Q$, $U$ and $V$ are normalized to the maximum
line-core depression (from the continuum level) of the Stokes $I$ profile
of the `red' absorption line. The weakness of the 
Stokes $V$ signal, which is caused by the longitudinal Zeeman effect, 
arises because the magnetic field vector is almost parallel
to the solar surface. Finally, note that, for the particular geometry
of this observation, an ambiguity of  $180^\circ$ is present
in the determination of the azimuth of the magnetic field vector.

\vfill
\eject

\begin{figure}
\psfig{figure=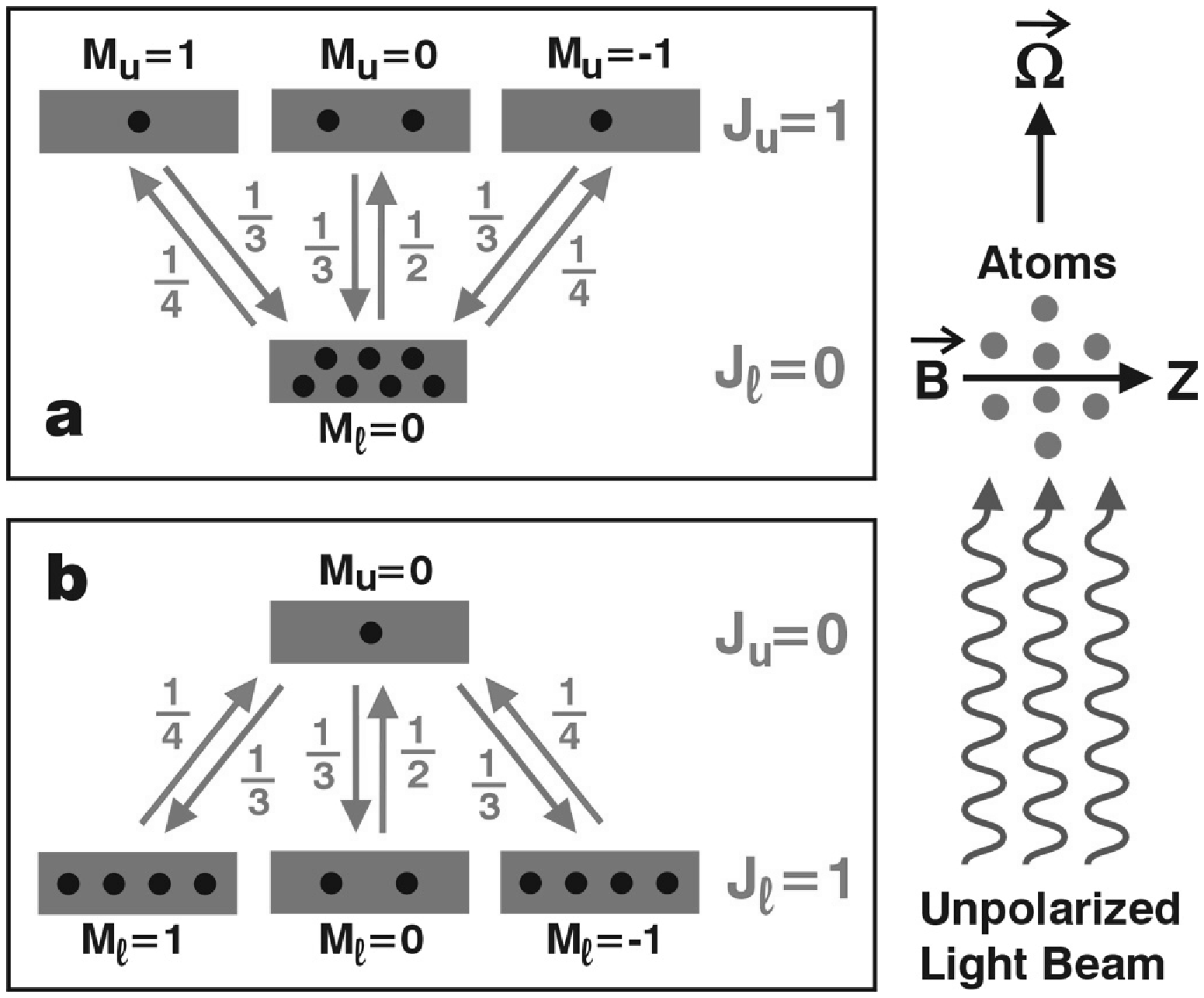}
\end{figure}

{\bf Figure 1}

\vfill
\eject

\begin{figure}
\psfig{figure=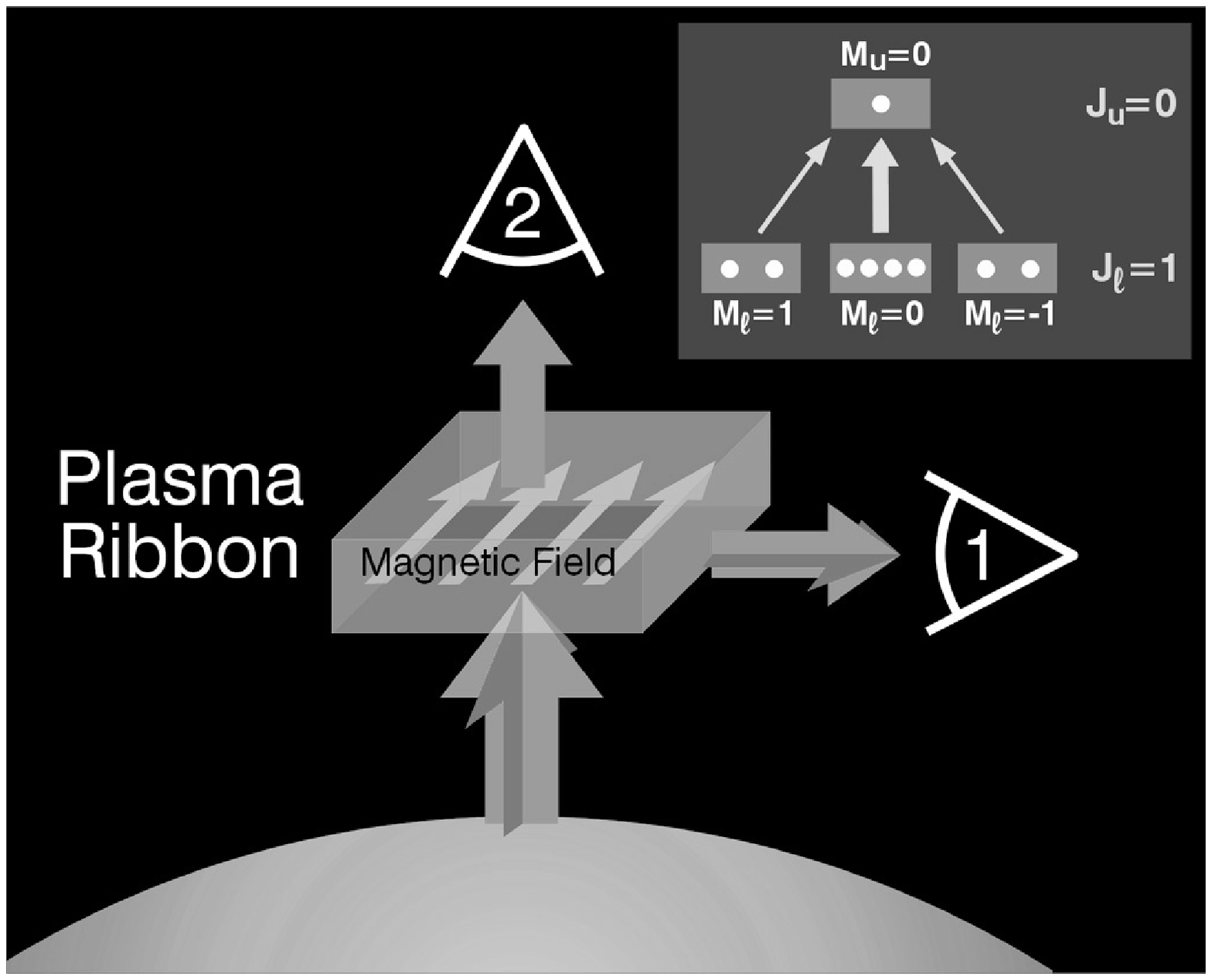}
\end{figure}

{\bf Figure 2}

\vfill
\eject

\begin{figure}
\psfig{figure=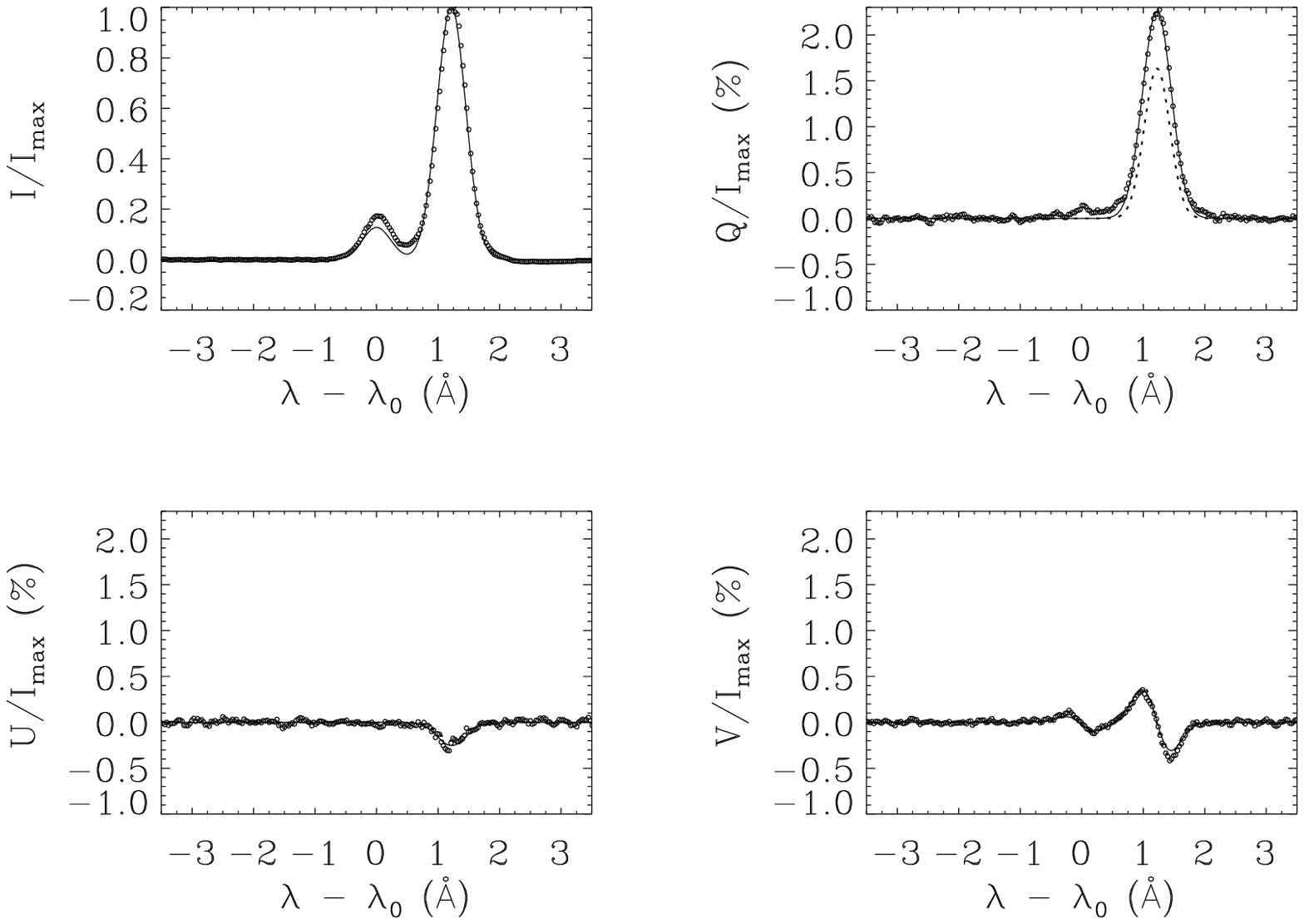,height=14cm,width=14cm}
\end{figure}

{\bf Figure 3}

\vfill
\eject

\begin{figure}
\psfig{figure=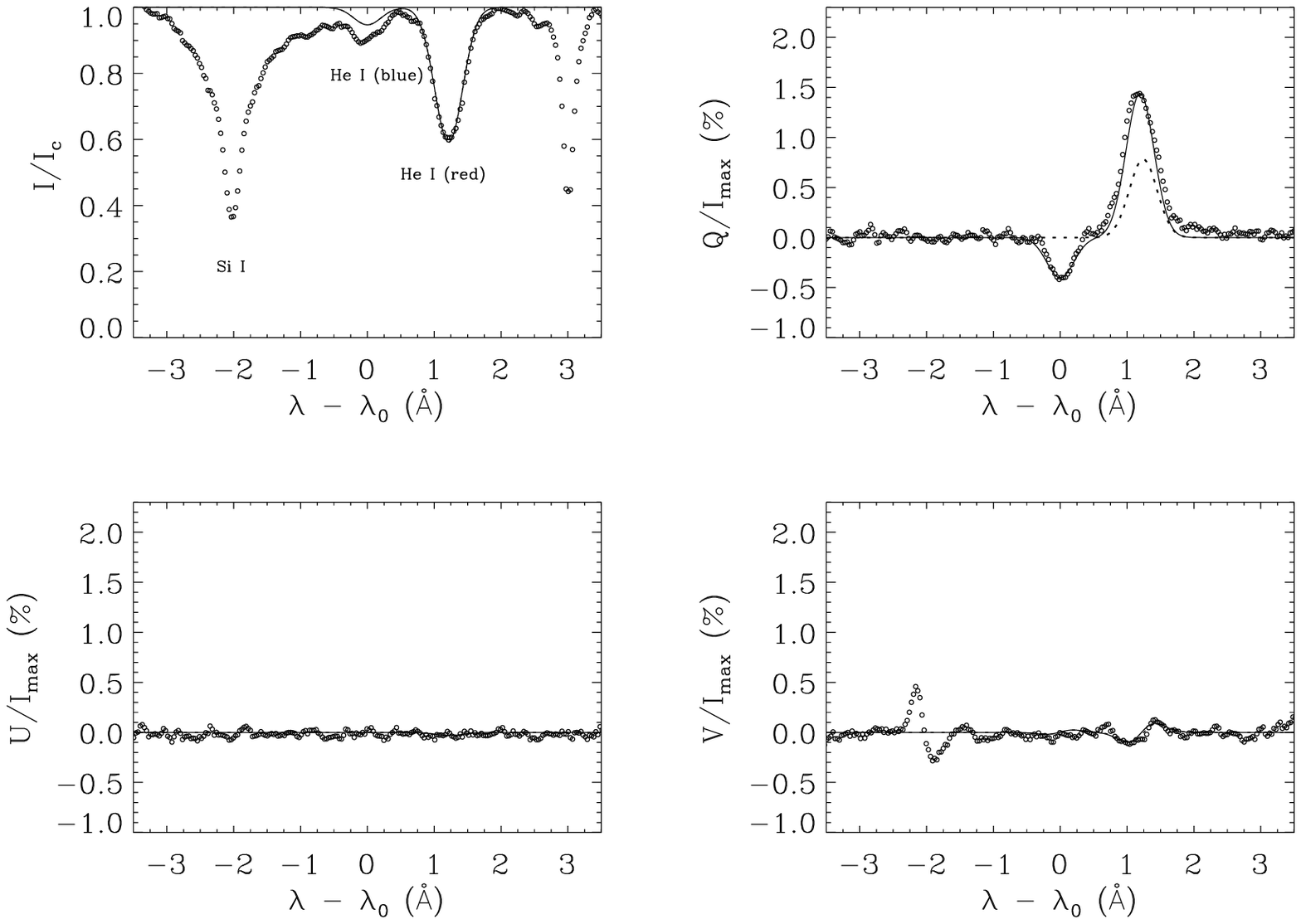,height=14cm,width=14cm}
\end{figure}

{\bf Figure 4}

\end{document}